\documentclass[aps,prl,twocolumn,superscriptaddress]{revtex4}
\usepackage{graphicx}
\usepackage{dcolumn}
\usepackage{bm}
\usepackage{url}

\begin{document}

\title{Conservation of high-flux backbone in alternate optimal and near-optimal flux distributions of metabolic networks}

\author{Areejit Samal}
\affiliation{Max Planck Institute for Mathematics in the Sciences,
Inselstr. 22, D-04103 Leipzig, Germany}
\email{samal@mis.mpg.de}

\begin{abstract}
Constraint-based flux balance analysis (FBA) has proven successful in predicting the flux distribution of metabolic networks in diverse environmental conditions. FBA finds one of the alternate optimal solutions that maximizes the biomass production rate. Almaas {\it et al} have shown that the flux distribution follows a power law, and it is possible to associate with most metabolites two reactions which maximally produce and consume a give metabolite, respectively.  This observation led to the concept of high-flux backbone (HFB) in metabolic networks. In previous work, the HFB has been computed using a particular optima obtained using FBA. In this paper, we investigate the conservation of HFB of a particular solution for a given medium across different alternate optima and near-optima in metabolic networks of {\it E. coli} and {\it S. cerevisiae}. Using flux variability analysis (FVA), we propose a method to determine reactions that are guaranteed to be in HFB regardless of alternate solutions. We find that the HFB of a particular optima is largely conserved across alternate optima in {\it E. coli}, while it is only moderately conserved in {\it S. cerevisiae}. However, the HFB of a particular near-optima shows a large variation across alternate near-optima in both organisms. We show that the conserved set of reactions in HFB across alternate near-optima has a large overlap with essential reactions and reactions which are both uniquely consuming (UC) and uniquely producing (UP). Our findings suggest that the structure of the metabolic network admits a high degree of redundancy and plasticity in near-optimal flow patterns enhancing system robustness for a given environmental condition.
\end{abstract}
\maketitle

\section*{Introduction}

The study of the structure and dynamics of complex biological networks is important for understanding the system-level behaviour of living organisms \citep{HHLM1999,BO2004}. The metabolic network is one such physicochemical system inside the cell which can be viewed as a complex network consisting of metabolites involved in enzyme catalyzed reactions. Recent advances in the development of high-throughput data collection techniques coupled with systematic analysis of fully sequenced genomes has led to the reconstruction of organism-specific metabolic networks \citep{FHTRP2009}. Although the list of metabolic reactions along with the involved stoichiometries is largely known for many organisms, we currently lack the knowledge of most kinetic rate constants and enzyme concentrations in these networks. In the absence of detailed knowledge of various kinetic parameters, constraint-based modelling techniques like Flux balance analysis (FBA), which utilizes primarily structural information of the network in the form of stoichiometry of various reactions, have proven useful in predicting the quantitative behaviour of the cellular metabolism under different environmental conditions \citep{VP1994b,EIP2001,KPE2003,PRP2004}.

FBA is computational technique which can be used to obtain a particular flux distribution for a given environmental condition that maximizes certain objective function which is usually taken to be the biomass production rate (see methods section for details). However, for large scale metabolic networks, it has been shown that there are a large number of different flux distributions or alternate solutions for a given environmental condition with the exact same value for the objective function \citep{LPDG2000,MS2003,RP2004}. Thus, FBA finds one of many possible alternate optimal solutions for a given environmental condition. Mixed integer linear programming (MILP) has been used to enumerate different alternate optimal flux distributions for a given environmental condition \citep{LPDG2000,RP2004}. However, for large scale metabolic networks, the number of alternate optimal solutions has been shown to be extremely large, and it is computationally infeasible using MILP to enumerate all possible alternate optima \citep{RP2004}. Although, it is not feasible to enumerate all alternate optima, flux variability analysis (FVA) \citep{MS2003} can be used to determine the minimum and maximum flux value of each reaction across the set of alternate optimal solutions for a given environmental condition. Specifically, using FVA, we can determine reactions that can have a nonzero flux in some alternate optima and the reactions whose flux vary across different alternate optima for a given environmental condition.

In the FBA approach, we obtain a particular solution by imposing the optimality criterion wherein a flux distribution satisfying the governing constraints along with maximal biomass production rate is predicted as the phenotype for a given environmental condition. However, the optimal flux distribution with maximal biomass production may not always explain the observed behaviour for a given environmental condition \citep{IEP2002,SVC2002,FJP2005,SKS2007,SPF2008}. Further, wet lab experiments have shown that the organism may initially grow suboptimally under certain media, i.e., the observed growth rate is less than the maximum value predicted by FBA, and then evolve to grow at the predicted optimal rate after several generations \citep{IEP2002,FJP2005}. It is plausible that there are multiple objectives that the cell is trying to optimize to different extent under different environments. Thus, the metabolic network may operate under suboptimal growth conditions under certain environmental conditions. Hence, it is important to study the suboptimal flux distributions, especially, near-optimal flux distributions in metabolic networks for different environmental conditions.

Using the optimal solution predicted by FBA, \citet{AKVOB2004} have shown that the distribution of reaction fluxes for a given environmental condition follows a power law in the {\it E. coli} metabolic network. The observed power law distribution implies an inhomogeneity in the reaction network usage with most reactions having small fluxes and a few reactions having high fluxes. \citet{AKVOB2004} showed that the heterogeneity of overall flux distribution is also valid to a large extent at the level of individual metabolites. Further, for most metabolites, they were able to identity two reactions that dominate the production and consumption of a metabolite, respectively. Using this property at the level of an individual metabolite, \citet{AKVOB2004} have proposed a simple algorithm to construct a subnetwork referred to as the ``high-flux backbone'' (HFB) which contains the locally maximal flow paths for a given environmental condition in the network . The HFB for a given environmental condition was shown to form a giant connected cluster with large overlap with classical pathways described in biochemical literature. It has also been suggested that the HFB can be used as a network reduction algorithm.

In this paper, we have studied the activity and flux variability of different reactions across alternate flux distributions for a given medium under optimal and suboptimal growth conditions in metabolic networks of {\it E. coli} and {\it S. cerevisiae}. \citet{AKVOB2004} have used a single solution obtained using FBA from the set of different alternate optima to compute the HFB of the metabolic network for a given medium. In this paper, we have investigated in detail the effect of alternate optima on the set of reactions in HFB computed using a single solution for a given medium in {\it E. coli} and {\it S. cerevisiae}. We have further investigated the effect of alternate suboptimal solutions corresponding to near-optimal growth condition on the set of reactions in HFB computed using a single suboptimal solution for a given medium in the two organisms.


\section*{Results and Discussion}

In this section, we present results from our study of the two metabolic networks: {\it E. coli} (version iJR904 \citep{RVSP2003}) and {\it S. cerevisiae} (version iND750 \citep{DHP2004}). The two networks can be downloaded from the BiGG database: {\it http://bigg.ucsd.edu/}. The {\it E. coli} metabolic network iJR904 contains 761 metabolites and 931 reactions and the {\it S. cerevisiae} metabolic network iND750 contains 1061 metabolites and 1149 reactions. For each organism, the two databases additionally provide a biomass production reaction which gives the relative contribution of key metabolites towards cellular biomass. We have investigated here the two metabolic networks under different environmental conditions corresponding to 89 and 43 aerobic minimal media for {\it E. coli} and {\it S. cerevisiae}, respectively. Using FBA, it has been predicted elsewhere that this set of 89 and 43 aerobic minimal media for {\it E. coli} and {\it S. cerevisiae}, respectively, considered here are the only aerobic minimal media that can support nonzero growth in the two networks \citep{SSGKRJ2006}.

\begin{figure*}
\centering
\includegraphics[width=11cm]{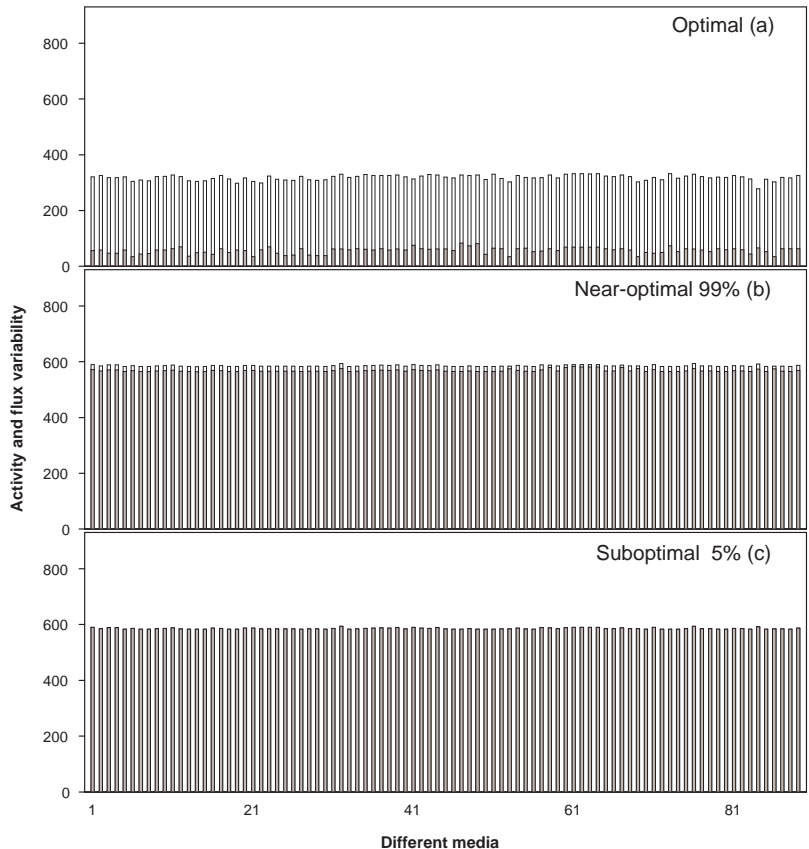}
\caption{Activity and flux variability of reactions under (a) optimal growth, (b) near-optimal growth ($\ge$ $99\%$ of the optimal rate) and (c) suboptimal growth ($\ge$ $5\%$ of the optimal rate) for 89 different minimal media in the {\it E. coli} metabolic network. Each bar corresponds to one of the 89 different minimal media. The height of the bar gives the number of reactions that can be possibly active and the height upto which the bar is filled with grey colour gives the number of reactions whose flux vary across alternate solutions for that medium.}
\label{ACTEC}
\end{figure*}

\begin{figure*}
\centering
\includegraphics[width=11cm]{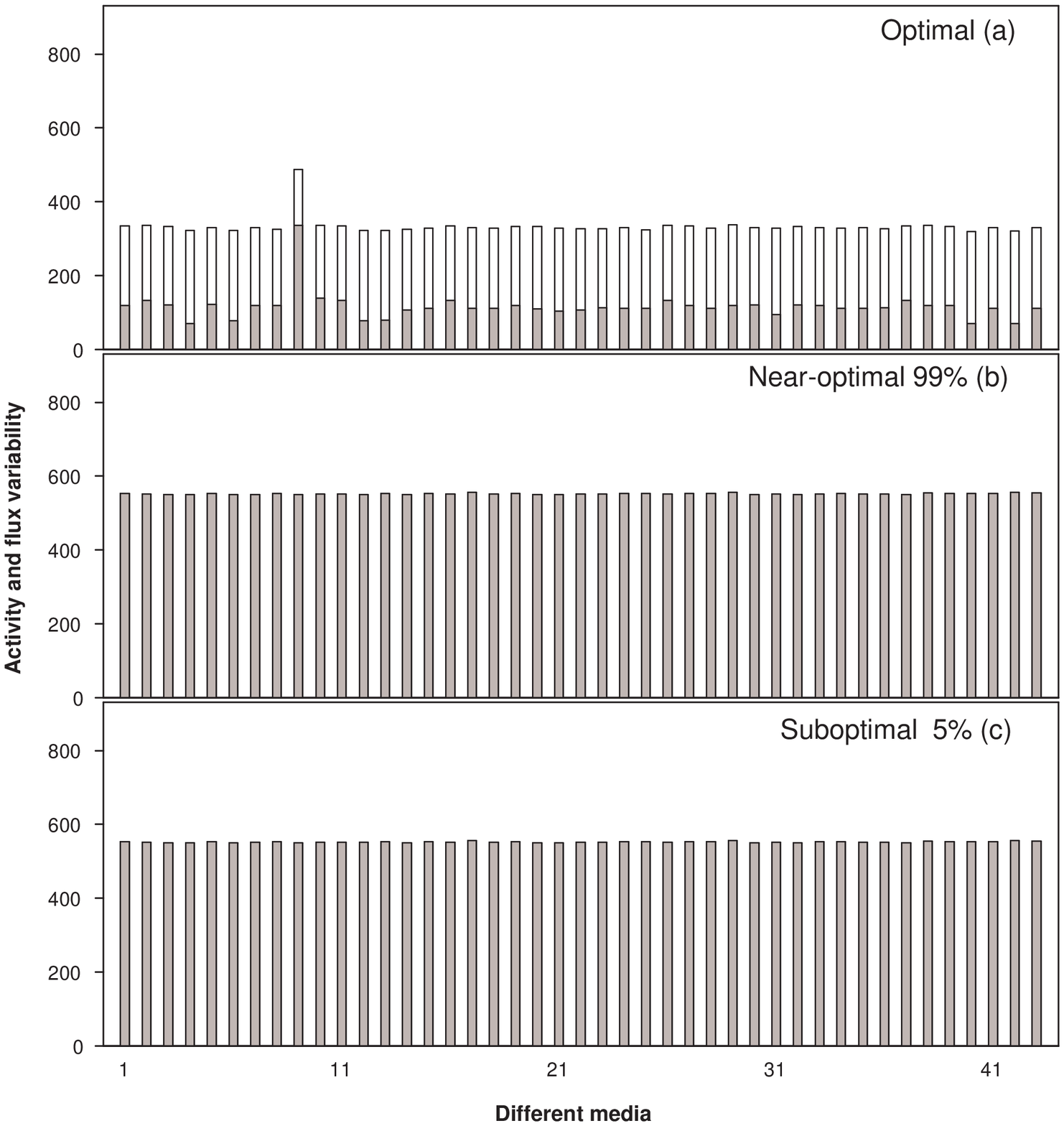}
\caption{Activity and flux variability of reactions under (a) optimal growth, (b) near-optimal growth ($\ge$ $99\%$ of the optimal rate) and (c) suboptimal growth ($\ge$ $5\%$ of the optimal rate) for 43 different minimal media in the {\it S. cerevisiae} metabolic network. Each bar corresponds to one of the 43 different minimal media. The height of the bar gives the number of reactions that can be possibly active and the height upto which the bar is filled with grey colour gives the number of reactions whose flux vary across alternate solutions for that medium.}
\label{ACTSC}
\end{figure*}

\subsection*{\bf Activity and flux variability of reactions under optimal and near-optimal conditions}

We can determine the maximum and minimum flux through each reaction that supports optimal growth in a minimal medium using FVA (see methods section for details). A reaction is considered to be possibly active in a medium across alternate optima, if the maximum flux value that supports optimal growth is nonzero. A reaction is considered to have a variable flux in a medium across alternate optima, if the difference between maximum and minimum flux value that supports optimal growth is nonzero.

Using FVA, we determined the number of reactions that can be possibly active and the reactions with variable flux across alternate optima for each of the 89 different minimal media in the {\it E. coli} metabolic network.  The number of reactions that can be possibly active in a medium varies between 278 and 332 reactions across the 89 minimal media. The number of reactions with variable flux in a medium varies between 34 and 83 reactions across the 89 minimal media. These results for the {\it E. coli} metabolic network are shown in Fig. \ref{ACTEC}(a). We find that $11-25\%$ of the reactions that can be possibly active in a given medium to have variable flux across alternate optima for that medium. The results obtained here for the {\it E. coli} metabolic network using FVA for 89 different aerobic minimal media are consistent with the results obtained earlier by \citet{RP2004} using a larger set of alternate optima for 136 different aerobic and anaerobic minimal media.

We also determined the number of reactions that can be possibly active and the reactions with variable flux across alternate optima for each of the 43 different minimal media in the {\it S. cerevisiae} metabolic network. These results for the {\it S. cerevisiae} metabolic network are shown in Fig. \ref{ACTSC}(a). We find that $22-69\%$ of the reactions that can be possibly active in a given medium to have variable flux across alternate optima for that medium in {\it S. cerevisiae}. For one of the 43 minimal media, i.e., arginine aerobic minimal medium, we find a much larger number of possibly active reactions and reactions with variable flux across alternate optima compared to other minimal media (cf. Fig. \ref{ACTSC}(a)). A closer look at the data for arginine minimal medium showed that many transport reactions involved in excretion of ammonia, guanine, urea and xanthine have a variable flux across alternate optima. This observation explains the larger flexibility under arginine minimal medium within the {\it S. cerevisiae} metabolic network to achieve optimal growth rate compared to other media. Even if we exclude the results for the arginine minimal medium, we find that the flux variability for a typical minimal medium in {\it S. cerevisiae} is greater than the flux variability for a typical minimal medium in {\it E. coli} under optimal growth condition. This finding suggests a more evolved structure of the metabolic network inside the eukaryote {\it S. cerevisiae} with greater redundancy compared to that of the prokaryote {\it E. coli}.

We investigated both metabolic networks under two suboptimal scenarios: (a) where the growth rate is $\ge 99\%$ of the optimal growth rate and (b) where the growth rate is $\ge 5\%$ of the optimal growth rate. The first of the two suboptimal scenarios will be referred to as `near-optimal condition' as we account for all flux distributions that allow growth at $\ge 99\%$ of the optimal rate. In the second suboptimal scenario, we account for all flux distributions which allow growth at $\ge 5\%$ of the optimal rate. We used FVA to determine the number of possibly active reactions and the reactions with variable flux in a medium across alternate suboptimal solutions for different minimal media. From here onwards, we refer to the set of alternate optimal flux distributions for a given minimal medium in short as `alternate optima'. For the first suboptimal scenario, we refer to the set of alternate suboptimal flux distributions which have near-optimal growth rate for a given minimal medium in short as `alternate near-optima'. For the second suboptimal scenario, we refer to the set of alternate suboptimal flux distributions which have growth rate $\ge 5\%$ of the optimal rate for a given minimal medium in short as `alternate suboptima'.

Using FVA, we determined the number of reactions that can be possibly active and the reactions with variable flux in the two suboptimal scenarios for each of the 89 different minimal media in the {\it E. coli} metabolic network. The number of reactions that can be possibly active in a given medium is obtained to be close to 600 in both suboptimal scenarios. Almost all reactions which can be possibly active have variable flux across alternate near-optima in the first suboptimal scenario. Further, all reactions which can be possibly active have variable flux across alternate suboptima in the second suboptimal scenario. These results for the {\it E. coli} metabolic network are shown in Fig. \ref{ACTEC}(b) and (c). We find that most reactions which can be possibly active in a given medium can have a variable flux across alternate solutions in near-optimal growth condition (cf. Fig. \ref{ACTEC}(b)). We found similar results for both suboptimal conditions in the {\it S. cerevisiae} metabolic network (cf. Fig. \ref{ACTSC}(b) and (c)). We emphasize that although the set of reactions with variable flux under both suboptimal scenarios is almost the same, the amount of flux variability of a reaction is larger in the second suboptimal scenario compared to the first near-optimal scenario.

By comparing the number of possibly active reactions in each minimal medium investigated in {\it E. coli} (cf. Fig. \ref{ACTEC}(a) and (b)) and {\it S. cerevisiae} (cf. Fig. \ref{ACTSC}(a) and (b)) under optimal and near-optimal conditions, we can clearly see that the number is significantly more in the near-optimal condition. Thus, a larger set of reactions can be recruited under near-optimal condition compared to strictly optimal condition. This suggests a high degree of flux plasticity in the metabolic network under near-optimal condition. In a recent paper, \citet{NGM2008} have shown that the maximization of a linear objective function, such as growth rate or ATP production, in FBA to obtain a particular flux distribution leads to significantly lower number of active reactions under optimal condition compared to typical non-optimal states. They attribute this massive spontaneous reaction silencing to the irreversibility of reactions and cascade of inactivity propagated by irreversible reactions in the network. Our results obtained using FVA add to the results obtained by \citet{NGM2008}.  We have shown here that, even in near-optimal growth conditions, the number of possibly active reactions and the reactions with variable flux across alternate near-optima is much larger compared to strictly optimal scenario.

We end this subsection by mentioning an important methodological detail. While implementing FVA, we need to account for reactions that are involved in futile internal cycles. An internal cycle or Type III extreme pathway is a set of internal reactions that can satisfy the steady state mass balance condition without involvement of any exchange flux \citep{BLQ2002,PFBP2002}. Such cycles are thermodynamically infeasible and need to be eliminated in any steady state analysis. We determined all internal cycles in the two metabolic networks studied here, and based on that, decided to constrain the flux of 8 reactions in {\it E. coli} and 16 reactions in {\it S. cerevisiae} to zero to avoid futile cycles in our simulations (see methods section for details).

\begin{figure*}
\centering
\includegraphics[width=11cm]{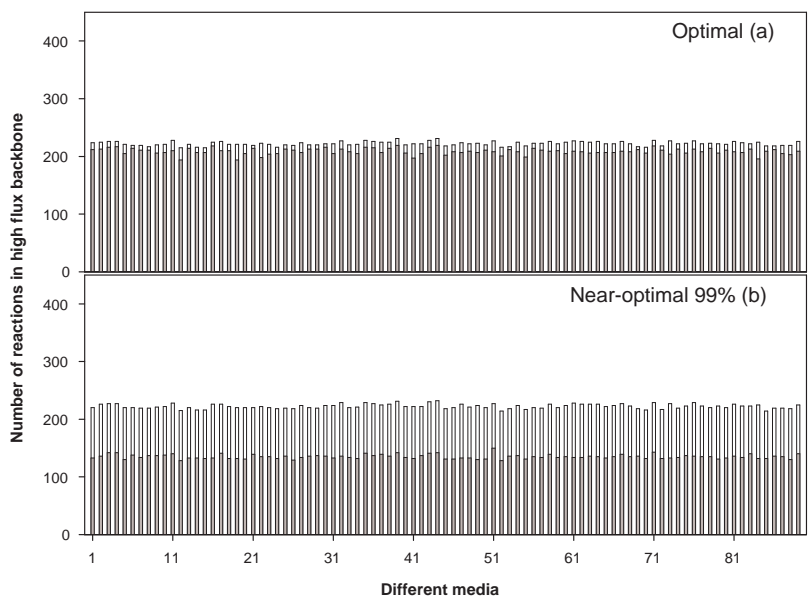}
\caption{Histogram of the average number of reactions in HFB under (a) optimal growth and (b) near-optimal growth for 89 different minimal media in the {\it E. coli} metabolic network. Each bar corresponds to one of the 89 different minimal media. The height of the bar gives the average number of reactions in the HFB computed using 100 alternate solutions obtained using MILP for each minimal medium. The height upto which the bar is filled with grey colour gives the number of reactions which are guaranteed to be in HFB of all alternate solutions for that medium.}
\label{HFBEC}
\end{figure*}

\begin{figure*}
\centering
\includegraphics[width=11cm]{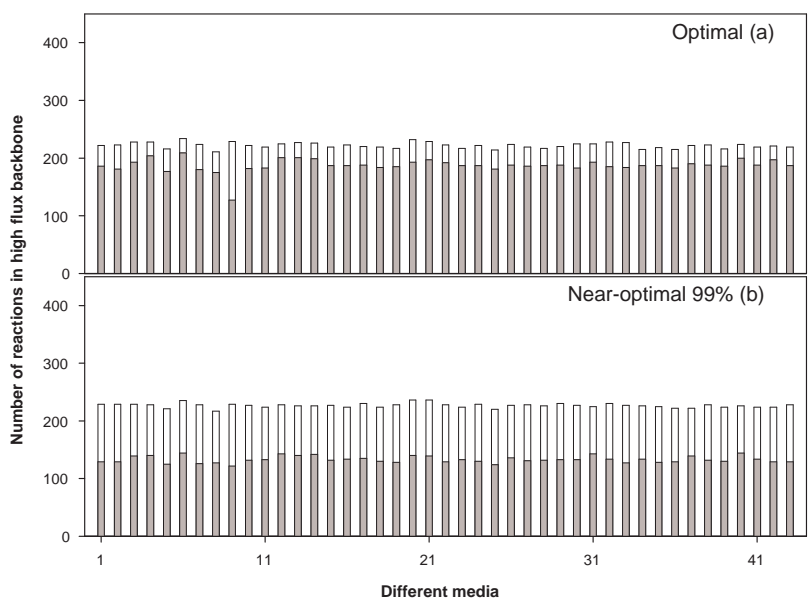}
\caption{Histogram of the average number of reactions in HFB under (a) optimal growth and (b) near-optimal growth for 43 different minimal media in the {\it S. cerevisiae} metabolic network. Each bar corresponds to one of the 43 different minimal media. The height of the bar gives the average number of reactions in the HFB computed using 100 alternate solutions obtained using MILP for each minimal medium. The height upto which the bar is filled with grey colour gives the number of reactions which are guaranteed to be in HFB of all alternate solutions for that medium.}
\label{HFBSC}
\end{figure*}

\subsection*{\bf High-flux backbone under optimal and near-optimal conditions}

In the previous subsection, we have shown that there is variation in the activity as well as flux value of reactions across the alternate optima for a given medium in both {\it E. coli} and {\it S. cerevisiae}. Further, we have shown that the number of possibly active reactions and reactions with variable flux across alternate near-optima is significantly higher in near-optimal condition compared to the strictly optimal case. In this subsection, we investigate the effect of alternate flux distributions on the set of reactions in the high-flux backbone (HFB) obtained using a particular solution in both optimal and near-optimal conditions.

Using mixed integer linear programming (MILP) \citep{LPDG2000,RP2004}, we obtained 100 alternate optima and 100 alternate near-optima for each of the 89 aerobic minimal media in {\it E. coli} and 43 aerobic minimal media in {\it S. cerevisiae} (see methods section for details). For each minimal medium investigated, we obtained the 100 HFB corresponding to 100 alternate optima. We then calculated the average number of reactions in HFB across the alternate optima for each medium. These results for the 89 minimal media in {\it E. coli} and 43 minimal media in {\it S. cerevisiae} under optimal growth conditions are shown in Fig. \ref{HFBEC}(a) and Fig. \ref{HFBSC}(a), respectively. Similarly, we obtained the average number of reactions in HFB across the alternate near-optima for each medium. These results are shown for {\it E. coli} and {\it S. cerevisiae} in Fig. \ref{HFBEC}(b) and Fig. \ref{HFBSC}(b), respectively.

In case of both {\it E. coli} and {\it S. cerevisiae}, for a given medium, we found the set of reactions in HFB corresponding to different alternate optima to differ from each other. This was also true for the the set of reactions in the HFB corresponding to different alternate near-optima. However, the number of alternate optima or alternate near-optima sampled here is too small, and the data cannot be used to determine the overlap among the reactions in HFB corresponding to different alternate solutions for a given medium. Thus, we devise a simple method using FVA to determine the set of reactions which are guaranteed to be in any HFB corresponding to different alternate solutions for a given medium (see methods section for details). We used this method to determine the set of reactions that are guaranteed to be in HFB corresponding to any alternate optima for a given medium in the two networks. We also obtained the set of reactions that are guaranteed to be in HFB corresponding to different alternate near-optima for a given medium in the two networks. These results are shown for the 89 minimal media in {\it E. coli} in Fig. \ref{HFBEC}(a) and (b) and for the 43 minimal media in {\it S. cerevisiae} in Fig. \ref{HFBSC}(a) and (b).

We find that at least $3-13\%$ of reactions in HFB corresponding to a particular optima for a given medium are not guaranteed to in HFB corresponding to some alternate optima for that medium in the {\it E. coli} metabolic network (cf. Fig. \ref{HFBEC}(a)). However, at least $34-42\%$ of reactions in HFB corresponding to a particular near-optima for a given medium are not guaranteed to be in HFB corresponding to some alternate near-optima for that medium  in {\it E. coli} (cf. Fig. \ref{HFBEC}(b)). Similarly, we find that at least $11-45\%$ of reactions in HFB corresponding to a particular optima for a given medium are not guaranteed to be in HFB corresponding to some alternate optima for that medium in the {\it S. cerevisiae} metabolic network (cf. Fig. \ref{HFBSC}(a)). Even if we discount the arginine minimal medium, which has an uncharacteristically larger flux variability in {\it S. cerevisiae} under optimal conditions (cf. Fig. \ref{ACTSC}(a)), still at least $11-20\%$ of reactions in HFB corresponding to a particular optima for a given medium are not guaranteed to be in HFB corresponding to some alternate optima for that medium. Further, at least $36-47\%$ of reactions in HFB corresponding to a particular near-optima are not guaranteed to be in HFB corresponding to some alternate near-optima for that medium in {\it S. cerevisiae} (cf. Fig. \ref{HFBSC}(b)).

Note that we have used the simplex method to solve the LP problem in order to sample different alternate optima and near-optima. Simplex method tries to return a particular solution or flux distribution from the set of alternate solutions that has least number of nonzero reaction fluxes in it \citep{Almaas2007}. Thus, our results obtained using MILP regarding the average number of reactions in HFB corresponding to a particular solution computed using 100 alternate solutions for a given medium may change with exhaustive sampling. However, our results obtained using FVA regarding the set of reactions that are guaranteed to be in HFB corresponding to any alternate optima or suboptima for a given medium is independent of sampling.

We have shown here that the set of reactions in HFB depends upon the particular solution chosen from the set of alternate optima to compute it in both {\it E. coli} and {\it S. cerevisiae}. Under optimal condition, only $3-11\%$ of the reactions in HFB of a particular solution are not guaranteed to be in HFB of some other alternate optima in {\it E. coli} which is relatively small. But the corresponding fraction in {\it S. cerevisiae} for optimal condition is much higher. Also, we have shown for near-optimal condition, greater than $34\%$ of the reactions in HFB of a particular near-optima are not guaranteed to be in HFB of some other alternate near-optima in both organisms. Thus, the set of reactions in HFB significantly varies under near-optimal conditions across alternate near-optima for a given medium in both organisms. These findings suggest that the flow patterns in metabolic networks for a given medium in near-optimal conditions is much more complex and degenerate compared to that under strictly optimal conditions.

Our findings also imply that even for a given medium the structure of the metabolic network has evolved so as to allow both structural and flux plasticity \citep{AOB2005} in the level of reaction usage to achieve the same optimal or near-optimal growth objective through multiple flux distributions. Structural plasticity associated with a given medium can result in alternate equally efficient pathways or metabolic routes to be recruited in different alternate solutions. Flux plasticity associated with a given medium can result in different value of reaction fluxes in different alternate solutions. Both structural and flux plasticity associated with alternate flux distributions for a given medium renders the set of reactions in HFB of a particular solution for that medium variable across alternate solutions. However, we still find that, under near-optimal condition, at least $53\%$ of the reactions in HFB of a particular solution are conserved across HFB of any alternate near-optima for a given medium in both organisms. In contrast, we have shown in the last subsection for near-optimal condition that, almost all reactions that can be possibly active for a given minimal medium have a variable flux across alternate near-optima for that medium.

\begin{figure*}
\centering
\includegraphics[width=11cm]{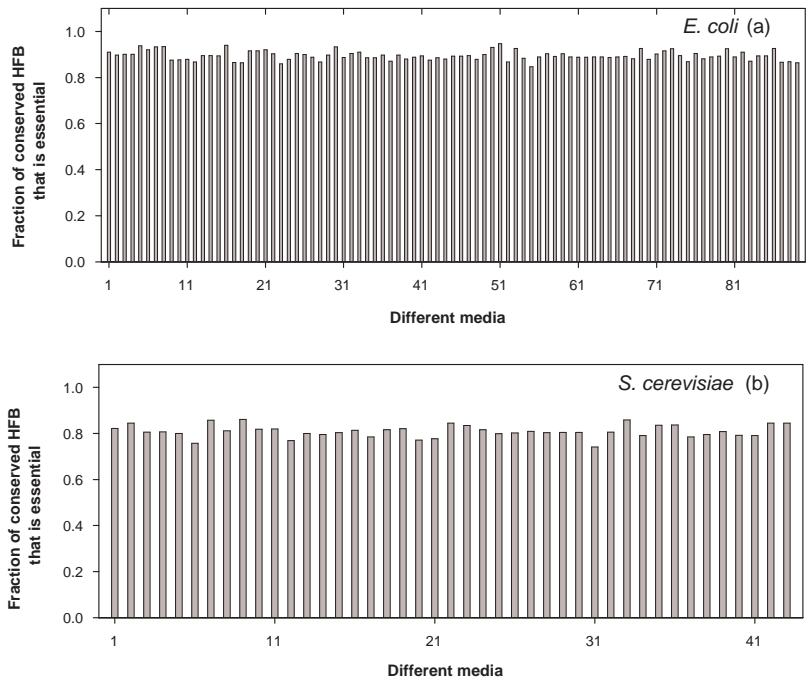}
\caption{Histogram of the fraction of reactions guaranteed to be in HFB of any alternate near-optima that are essential for (a) 89 minimal media in {\it E. coli} and (b) 43 minimal media in {\it S. cerevisiae}.}
\label{Essential}
\end{figure*}

We now shed light on why at least $53\%$ of the reactions in HFB of a particular solution are conserved across HFB of any alternate near-optima for a given medium when almost all reactions that can be possibly active for that medium show variation in their flux value. Since, the set of conserved reactions across HFB of any alternate near-optima are always active for a given medium, it is natural to expect this set of reactions to have a high overlap with essential reactions for growth in that medium. We find that on average $90\%$ of reactions guaranteed to be in HFB of alternate near-optima are essential for a given medium in {\it E. coli} while the corresponding fraction in {\it S. cerevisiae} is $80\%$ (cf. Fig. \ref{Essential}). Thus, the conserved part of HFB across alternate solutions for a given medium has high density of essential reactions.

\begin{figure*}
\centering
\includegraphics[width=11cm]{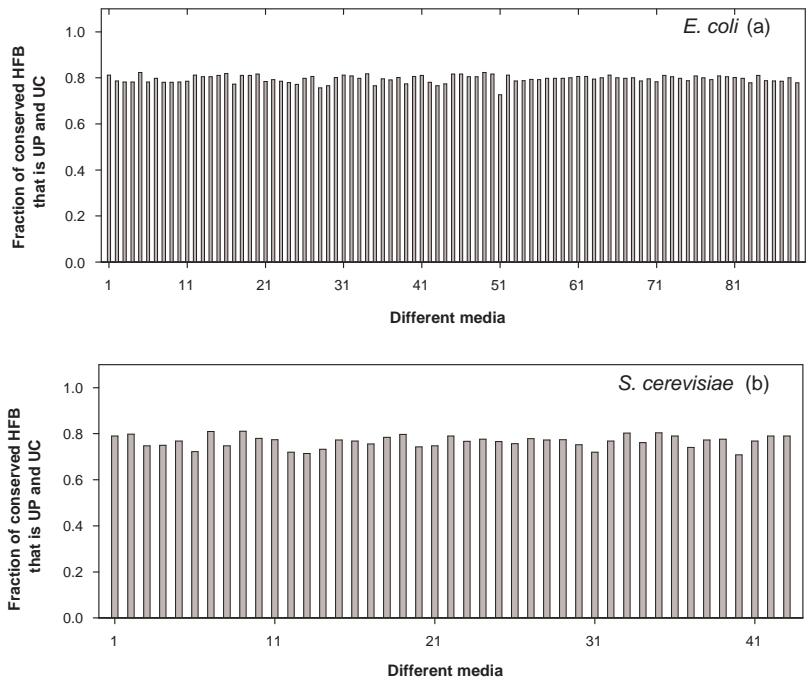}
\caption{Histogram of the fraction of reactions guaranteed to be in HFB of any alternate near-optima that satisfy both UP and UC property for (a) 89 minimal media in {\it E. coli} and (b) 43 minimal media in {\it S. cerevisiae}.}
\label{UPUC}
\end{figure*}

It has been shown elsewhere that almost all essential reactions in the metabolic network get explained by their association with low degree uniquely produced (UP) or uniquely consumed (UC) metabolite \citep{SSGKRJ2006,PCGF2005}. Notice that if a given reaction consumes a UC metabolite and produces a UP metabolite, and the reaction has a nonzero flux value across all alternate near-optima for a given medium, then it is guaranteed to be in HFB of any alternate near-optima. We find that on average $80\%$ of reactions guaranteed to be in HFB of alternate near-optima consume a UC metabolite and produce a UP metabolite for a given medium in {\it E. coli} while the corresponding fraction for {\it S. cerevisiae} is $77\%$ (cf. Fig. \ref{UPUC}). Thus, the conserved part of HFB across alternate solutions for a given medium has high density of reactions that satisfy both UP and UC property. In determining UP and UC metabolites for a given medium, we removed the blocked reactions \citep{SS1991,BNSM2004} for that medium as outlined in \citet{SSGKRJ2006}.

The above mentioned results show that most reactions that are conserved in HFB regardless of alternate near-optima for a given medium get explained by the absence of multiple paths to carry out certain processes in the metabolic network. Our findings also suggests that, in most cases, if a metabolite has multiple reactions producing it and multiple reactions consuming it, we cannot associate the metabolite with a single reaction that produces it maximally across all alternate near-optima and a single reaction that consumes it maximally across all alternate near-optima. These results show that, under near-optimal conditions, the metabolic network of {\it E. coli} and {\it S. cerevisiae} exhibit a high degree of flux plasticity for a given minimal medium contributing towards the robustness of the system.


\section*{Conclusions}

In this paper, we have investigated the effect of alternate optima and near-optima for a given medium on the set of reactions in the high-flux backbone (HFB) computed using a particular solution for that medium in the {\it E. coli} and {\it S. cerevisiae} metabolic network. Using flux variability analysis (FVA), we have shown here that the number of reactions that can be possibly active and reactions with variable flux across alternate solutions for a given medium under strictly optimal condition is significantly less than that in near-optimal condition for both metabolic networks. Since, the set of reactions in the HFB of a particular solution may vary across different alternate solutions for a given medium, we devised a simple method using FVA to determine the set of reactions which are guaranteed to be in HFB of any alternate solution for a given medium. We find that the set of reactions in HFB of particular optima for a given medium shows small variation across alternate optima for that medium in {\it E. coli} and moderate variation in {\it S. cerevisiae}. However, this variation is significantly large in both organisms when we consider the alternate near-optima for a given medium. We show that the majority of reactions that are guaranteed to be in HFB of any alternate near-optima for a given medium are also essential for growth in that medium. We also show that majority of reactions in the conserved HFB set across alternate near-optima for a given medium also satisfy both the uniquely consuming (UC) and uniquely producing (UP) property. Our results show that the structure of the metabolic network allows a high degree of structural and flux plasticity within a given environmental condition under near-optimal conditions which enhances the robustness of metabolic networks in both organisms.

\section*{Methods}

\subsection*{\bf Flux balance analysis}

Flux balance analysis (FBA) is a computational modelling technique which can be used to obtain a prediction for the fluxes of all reactions in the metabolic network and growth rate of the organism \citep{VP1994b,KPE2003,PRP2004}. The main assumptions made in the technique of FBA are: (a) the network reaches a steady state for any given environmental condition wherein the concentration of all internal metabolites and velocities of all reactions are constant, and (b) the cell tries to adjust its intracellular machinery or reaction fluxes so as to maximize its growth rate or biomass production. The key requirement for implementing FBA is the information about the stoichiometry of all metabolic reactions known to occur in an organism. The complete list of reactions along with their stoichiometry is encapsulated in the stoichiometric matrix ${\bf S}$ of dimensions $m$ $\times$ $n$ where $m$ denotes the number of metabolites and $n$ denotes the number of reactions. In any metabolic steady state, each metabolite achieves a dynamic mass balance and the vector ${\bf v}$ = ($v_1$,$v_2$,\ldots,$v_n$)$^T$ representing the fluxes through each reaction must satisfy the equation
\begin{equation}
\label{steadystate}
{\bf S}.{\bf v} = 0,
\end{equation}
so as to not violate mass conservation. Eq. \ref{steadystate} gives us a system of linear equations relating various fluxes in the network. In addition to linear stoichiometric constraints, there exist thermodynamic constraints that may render certain reactions irreversible and flux capacity constraints which impose limitations on maximum flux through certain reactions. These additional constraints can be used to further limit the space of allowable fluxes given by Eq. \ref{steadystate}. To obtain a particular solution from the space of allowable fluxes, linear programming (LP) is used to find a flux distribution in the allowable space that gives the optimal value for an objective function. The most commonly used objective function is the maximization of biomass production rate. The LP formulation of the FBA approach is as follows:
\begin{eqnarray}
\label{fba}
& \mbox{Maximize} & Z={\bf c^{T}}{\bf v} \nonumber \\
& \mbox{such that} & {\bf S}.{\bf v} = 0 \nonumber \\
& & \alpha\ \le\ v_i\ \le\ \beta
\end{eqnarray}
where vector ${\bf c}$ corresponds to the objective function, $\alpha$ and $\beta$ are the lower and upper limits of individual flux values.

\subsection*{\bf Mixed integer linear programming}

For most large scale metabolic networks, FBA finds one of many possible alternate optimal flux distributions that have the same objective value for a given environmental condition \citep{LPDG2000,MS2003,RP2004}. We can use mixed integer linear programming (MILP) to enumerate different alternate optima. We have used here the recursive MILP algorithm proposed by \citet{RP2004} to enumerate different alternate optima. In this method, in addition to constraints in the LP formulation of FBA given by Eq. \ref{fba}, the following new constraints are imposed to obtain different solutions:
\begin{eqnarray}
\label{milp}
& & \sum_{i\in NZ^{J-1}} y_i \ge 1 \nonumber \\
& & \sum_{i\in NZ^{J}} w_i \ge |NZ^{k}|-1, \quad k=1 \ldots J-1 \nonumber \\
& & y_i + w_i \le 1, \quad \forall i \nonumber \\
& & \alpha \cdot w_i \le\ v_j \le \beta \cdot w_i \quad \forall i
\end{eqnarray}
In this method each reaction $i$ in the network is associated with two binary variables $y_i$ and $w_i$. $NZ^{J}$ represents the set of reactions with nonzero flux value in the solution at iteration $J$. Given the constraints in Eq. \ref{milp}, if the binary variable $y_i$=1 then $w_i$=0, which forces both upper and lower bounds for reaction $i$ to be zero. Thus, setting $y_i$=1 forces the reaction $i$ to be removed from the basis of the next iteration. In this recursive algorithm, at each iteration $J$, at least one of the nonzero fluxes of the previous solution ($NZ^{J-1}$) is removed from the basis of the current iteration. The MILP problem given by Eq. \ref{fba} and \ref{milp} is then solved again to generate a different alternate solution.

\subsection*{\bf Flux variability analysis}

The number of alternate optima in large scale metabolic networks is extremely large, and it can be computationally infeasible to enumerate all alternate optima. Flux variability analysis (FVA) \citep{MS2003} is an alternative technique which can be used to determine the flux variability associated with each reaction in the complete set of alternate optima for a given medium. Initially, the maximum value of the objective function, i.e., $Z_{obj}$, is determined. This is followed by solving two LP problems for each reaction $i$ in the network. The LP formulation of the FVA approach is as follows:
\begin{eqnarray}
\label{fvamax}
\mbox{Case 1:} & & \nonumber \\
& \mbox{Maximize} & v_i \nonumber \\
& \mbox{such that} & {\bf S}.{\bf v} = 0 \nonumber \\
& & {\bf c^{T}}{\bf v}=Z_{obj} \nonumber \\
& & \alpha\ \le\ v_i \le\ \beta \quad \mbox{for}\ i=1 \ldots n
\end{eqnarray}
\begin{eqnarray}
\label{fvamin}
\mbox{Case 2:} & & \nonumber \\
& \mbox{Minimize} & v_i \nonumber \\
& \mbox{such that} & {\bf S}.{\bf v} = 0 \nonumber \\
& & {\bf c^{T}}{\bf v}=Z_{obj} \nonumber \\
& & \alpha\ \le\ v_i\ \le\ \beta \quad \mbox{for}\ i=1 \ldots n
\end{eqnarray}
In the first case, the maximum flux through each reaction that supports optimal growth rate $Z_{obj}$ is determined. In the second case, the minimum flux through each reaction that supports optimal growth rate $Z_{obj}$ is determined. The difference of maximum and minimum value that supports optimal growth rate for each reaction gives the flux variability of that reaction. A slight modification of the LP problem given by Eq. \ref{fvamax} and \ref{fvamin} is able to generate the flux variability under suboptimal conditions. The suboptimal condition corresponding to near-optimality studied here requires the following change in Eq. \ref{fvamax} and \ref{fvamin}: Replace the condition ${\bf c^{T}}$${\bf v}$$=$$Z_{obj}$ with ${\bf c^{T}}$${\bf v}$$\ge$$0.99$$Z_{obj}$.

\subsection*{\bf Reactions involved in futile cycles}

We used a recently published algorithm by \citet{WW2008} to compute the reactions in {\it E. coli} and {\it S. cerevisiae} metabolic network that are involved in internal cycles \citep{BLQ2002,PFBP2002} of size $\ge$ 3. The number of reactions involved in internal cycles of size $\ge$ 3 in {\it E. coli} and {\it S. cerevisiae} was obtained to be 22 and 60, respectively. Using METATOOL version $5.0$ \citep{VS2006}, we obtained a total of 9 and 28 internal cycles of size $\ge$ 3 in the metabolic network of {\it E. coli} and {\it S. cerevisiae}, respectively. In order to avoid these internal cycles, we followed the standard procedure of constraining at least one reaction flux in each internal cycle to zero. By examining the different internal cycles of size $\ge$ 3, we decided to constrain the flux of following reactions: {\small ABUTt2, ACCOAL, ADK1, GLUT4, PROt4, SERt4, THRt4 and VPAMT} in the {\it E. coli} metabolic network iJR904 \citep{RVSP2003} and the flux of following reactions: {\small ASPTAm, ACONTm, DCMPDA, FRDcm, GHMT2r, GLUT5m, MALt2r, NDPK1, NDPK8, NDPK9, PHCDm, SHSL1, SUCCtm, SUCCFUMtm, SUCD2\_u6m and UGLT} in the {\it S. cerevisiae} metabolic network iND750 \citep{DHP2004} to zero to avoid futile cycles. The abbreviations of the reactions in the two metabolic networks mentioned above are same as in the databases iJR904 and iND750, respectively.

\subsection*{\bf High-flux backbone}

The method to determine high-flux backbone (HFB) for a given medium using a particular solution is as follows \citep{AKVOB2004}:
\begin{itemize}
\item[(a)] For each metabolite in the network, remove all reactions except the reaction that produces the metabolite with largest flux and the reaction that consumes the metabolite with largest flux. The metabolites that are either not produced (consumed) are not taken into account.
\item[(b)] If metabolite $A$ is an educt and metabolite $B$ is a product of reaction $R$, and further $R$ maximally consumes $A$ and maximally produces $B$, then reaction $R$ is part of the HFB.
\end{itemize}
Since, the set of reactions in HFB of a particular solution may change if it was obtained from a different alternate solution, we here propose a simple method using FVA to determine the set of reactions that are guaranteed to be in HFB of all alternate solutions for a given medium. Our method is as follows:
\begin{itemize}
\item[(a)] Using FVA, we obtain the minimum and maximum flux of each reaction $R$ in the network that can support optimal growth for a given minimal medium.
\item[(b)] We designate reaction $R$ as producing (consuming) metabolite $A$ with largest flux across all alternate optima, if the minimum flux for $R$ obtained using FVA is greater than the maximum flux for each reaction other than $R$ that produces (consumes) $A$.
\item[(c)] For each metabolite in the network, remove all reactions except the reaction that produces the metabolite with largest flux and the reaction that consumes the metabolite with largest flux across all alternate optima.
\item[(d)] If metabolite $A$ is an educt and metabolite $B$ is a product of reaction $R$, and further $R$ maximally consumes $A$ and maximally produces $B$ across all alternate optima, then reaction $R$ is contained in HFB of all alternate optima.
\end{itemize}
Our method described here for alternate optimal solutions can also be used to determine the set of reactions that are guaranteed to be in HFB of all alternate near-optimal solutions.

\subsection*{Acknowledgement}

The author wishes to thank Andreas Wagner and Thimo Rohlf for discussions, and Pierre-Yves Bourguignon for his feedback on the manuscript. He gratefully acknowledges a postdoctoral fellowship from the Max Planck Institute for Mathematics in the Sciences, Leipzig.



\end{document}